\documentclass[prd,showpacs,nofootinbib,preprintnumbers,preprint,floatfix]{revtex4}

\usepackage{amsmath}
\usepackage{amssymb}
\usepackage{dsfont}
\usepackage{multirow}
\usepackage{makecell}
\usepackage{bm}

\makeatother

\def\e{{\rm e}}
\def\be{\begin{equation}}
\def\ee{\end{equation}}

\begin{document}

\title{Supergravity p-branes with scalar charge}

\author{I. \surname{Bogush}$^{a}$}
\email{igbogush@gmail.com}
\author{D. \surname{Gal'tsov}$^{a,b}$}
\email{galtsov@phys.msu.ru} \affiliation{$^{a}$Faculty of Physics, Moscow State University, 119899, Moscow, Russia\\$^{a}$ Kazan Federal University, 420008 Kazan, Russia}

\begin{abstract}
Standard dilatonic supergravity $p$-branes have scalar charges that are not independent parameters, but are determined by the brane tension and Page charges. This feature can be traced to the no-hair theorem in the four-dimensional Einstein-scalar gravity, implying that more general solutions with independent scalar charges can have naked singularities. Since singular branes are also of interest as tentative classical counterparts of unstable tachyonic branes and/or brane-antibrane systems, it is worth investigating branes with independent scalar charges in more detail. Here we study singular branes associated with the Fisher-Janis-Newman-Winicour solution of four-dimensional gravity. In the case of codimension three, we also construct singular branes endowed with a Zipoy-Voorhees-type oblateness parameter. It is expected that such branes will not be supersymmetric in the string theory. We demonstrate this in the special case of NS5-branes of type II theory. We analyze geodesics and test scalar perturbations of new solutions focusing on possible quantum healing of classical singularities. 
 
\end{abstract}


\maketitle

\section{Introduction} \label{sec:introduction}
Supergravity dilatonic $p$-branes are solutions of the Einstein equations with dilaton and antisymmetric forms \cite{Duff:1993ye,Lu:1995yn,Stelle:1996tz}. As is well-known, a single electrically or magnetically charged asymptotically flat brane with isotropic transverse space  satisfying the cosmic censorship condition is determined by three parameters: tension, Page charge, and asymptotic dilaton value~\cite{Galtsov:2004puk}. 
The scalar dilaton charge is also present, but it is of a secondary nature and is not a free parameter in accordance with the well-known no-hair theorem in the case $p=0$. Black dilatonic branes have two horizons, the internal one being singular. So their extremal limit has a singular even horizon, though still exhibits supersymmetry. It turns out, however, that geodesic distance to the singularity is infinite, so the extremal horizon is not observed as naked singularity  \cite{Duff:1991pe}.

More general solutions were  found by direct integration of Einstein equations (see, e.g.,\cite{Zhou:1999nm}), but these were shown to contain  naked singularities~\cite{Galtsov:2004puk}. Singular branes  with extra parameters can also be of interest as tentative classical  counterparts of tachyonic branes or brane-antibrane systems~\cite{Lu:2004dp,Lu:2004ms,Kobayashi:2004ay,Galtsov:2005thm,Asakawa:2005vb}. 
However, the general structure of singular branes and their relationship to singular solutions of general relativity do not seem to have been systematically investigated. 
As a step in this direction,
we build and explore singular branes related to the famous Fisher-Janis-Newman-Winicour (FJNW) solution of Einstein's four-dimensional gravity minimally coupled to a massless scalar field.
Recall that the FJNW solution, first found by Fisher in 1948 \cite{Fisher:1948yn}, was rediscovered in various forms in  Refs. \cite{Buchdahl:1959nk,Janis:1968kn, Winicour:1968zz, Wyman:1981bd}, the equivalence  of which is shown in \cite{Virbhadra:1997ie,Bhadra:2001fx}. 
This solution has a strong curvature singularity on the (would be) event horizon \cite{Chase:1970,Virbhadra:1995iy, Abdolrahimi:2009dc}. It can be extended to higher dimensions \cite{Sokolowski:1986vq} and generalized to rotating solutions \cite{Bogush:2020lkp,Astorino:2014mda,Chauvineau:2018zjy}.

Recently, FJNW solution has become popular in four dimensions, as modeling deviations from the standard paradigm of the black hole physics in scalar-tensor theories \cite{Virbhadra:2002ju, Gyulchev:2008ff, Crisford:2017zpi, DeAndrea:2014ova}. 
Geodesics, charged particle trajectories, accretion disks in FJNW and associated backgrounds were studied 
\cite{Zhou:2014jja, Chowdhury:2011aa,Babar:2015kaa}. 
Some predictions were formulated for observations that should be taken into account in astrophysics in search of new physics. 
Additional interest in FJNW stems from the fact that its counterpart in modified gravity models such as Horndesky or DHOST may provide a non-singular solution to these theories~\cite{BenAchour:2019fdf,Galtsov:2020jnu}.
  
The FJNW solution with a certain value of the scalar charge  can be recognized as the solution of the Einstein-Maxwell-dilaton (EMD) theory in the limit of a vanishing electromagnetic field. Mathematically, this connection can be attributed to the existence of a three-dimensional sigma model, the target space of which contains a subspace corresponding to Einstein's minimal scalar theory. This fact can also be used as a generating tool. The  isometry transformations of the target space can be used to generate charged black holes with singular horizon starting with FJNW seed in four dimensions  \cite{Bogush:2020lkp}.

A similar generation technique has been proposed  \cite{Galtsov:1998mhf} for multidimensional gravity-antisymmetric form actions, and here we use it to generate $p$-brane solutions  with a scalar charge as an independent parameter. It is achieved via the application of the generalized Harrison transformations~\cite{Harrison:1968} to a Fisher-related seed, which has no electric and/or suitable uncharged seed. The new solution opens a way to contact the standard BPS dilatonic branes from a new perspective.   For a particular value of the dilaton charge, the new family reduces to the standard class of regular $p$-branes. This family contains, in particular, the NS5-brane from type IIA/B supergravities.
By explicitly checking the Killing spinor equations, we prove that the extended family  contains  no new supersymmetric solutions, except for the well-known extremal dilaton class.

In the particular case of three-dimensional transverse space, we also derive more general $p$-brane solutions endowed, apart from an independent scalar charge, with  an additional oblateness parameter of Zipoy-Voorhees type \cite{Zipoy:1966, Voorhees:1971wh}. This is possible due to some special symmetry of the Weyl-class solutions with the scalar charge.
 
We discuss  behavior of geodesics near singularities, identifying conditions when the latter are reached in finite and/or infinite time. We also consider  behavior of the test scalar field near singularities, investigating whether they can be regarded as unobservable in quantum theory. 

The plan of the paper is as follows. In Sec.II we briefly recall the 
derivation of the Harrison transformation for branes. This technique is then applied to construct Fisher branes in arbitrary spacetime dimensions in Sec.III. In the same section, we classify new solutions according to their singularity structure, discuss their supersymmetry, and show how to add Zipoy-Voorhees deformation to the solution. The next Sec. IV and V are devoted to the study of geodesics and the test scalar field near the singularities.

\section{Harrison transformations for branes} \label{sec:generation}
Here we briefly  recall the generating technique for $p$-branes \cite{Galtsov:1998mhf}. Consider Einstein gravity with  dilaton  and antisymmetric form in $D$-dimensional spacetime
\begin{equation}\label{eq:action}
    S =\frac1{2\kappa_D^2} \int d^D x \sqrt{-G} \left(
        R - \frac{1}{2} \left(\nabla\phi\right)^2 - \frac{\exp{(-\alpha \phi)}}{2(n+1)!}F^2_{(n+1)}
    \right),
\end{equation}
where $G_{MN}$ is a $D$-dimensional spacetime metric, $F_{(n+1)}=dA_{(n)}$ is an antisymmetric $n+1$-form, $\phi$ is a dilaton field, and $\alpha$ is the dilaton coupling constant, whose value is prescribed in a concrete theory. The antisymmetric form is squared with the weight one $F^2 = F_{M_1\ldots M_{n+1}}F^{M_1\ldots M_{n+1}}$.
The corresponding equations of motion are
\begin{subequations}\label{eq:eoms}
\begin{equation}\label{eq:eom_gravity}
    R_{MN} - \frac{1}{2} G_{MN} R = \e^{-\alpha\phi} T_{MN}^{(F)} + T_{MN}^{(\phi)},
\end{equation}
\begin{equation}\label{eq:eom_maxwell}
    \partial_M \left(
        \e^{-\alpha\phi} \sqrt{-G} F^{M M_1 \ldots M_n}_{(n+1)}
    \right) = 0,
\end{equation}
\begin{equation}\label{eq:eom_scalar}
    \frac{1}{\sqrt{-G}} \partial_M\left(
        \sqrt{-G} G^{MN} \partial_N \phi
    \right)
    +
    \frac{\alpha}{2(n+1)!} \e^{-\alpha\phi}F_{(n+1)}^2 = 0,
\end{equation}
\end{subequations}
where the energy-momentum tensors for the antisymmetric form and the dilaton read
\begin{subequations}
\begin{equation} \label{eq:energy_momentum_F}
    T_{MN}^{(F)} = \frac{1}{2n!} \left(
        F_{M M_1 \ldots M_n} F_N^{\;\;M_1 \ldots M_n}
    \right)
    - \frac{1}{4(n+1)!}G_{MN}F_{(n+1)}^2 ,
\end{equation}
\begin{equation} \label{eq:energy_momentum_phi}
    T_{MN}^{(\phi)} = \frac{1}{2}\left(
        \partial_M \phi \partial_N \phi - \frac{1}{2} G_{MN} \partial_L \phi \partial^L \phi
    \right).
\end{equation}
\end{subequations}
We assume that the brane is translation invariant in  space and time, which implies the existence of $d$ commuting Killing vectors, one of which is timelike. In adapted coordinates, the metric of the $D$-dimensional spacetime can be written in the form 
\begin{equation}\label{eq:metric_ansatz}
    ds^2 =
     g_{\mu\nu}(x) dy^\mu dy^\nu
    +\left(\sqrt{-g}\right)^{-2/s} h_{\alpha\beta}(x)dx^\alpha dx^\beta,
\end{equation}
where $g_{\mu\nu}$ and $h_{\alpha\beta}$ are metrics on the $d$-dimensional brane world-volume and the $s+2$-dimensional transverse space respectively, so that $D=d+s+2$.
Both metrics depend on the transverse coordinates only, and their indices vary as $\mu,\nu=0,\ldots,d-1$ and $\alpha,\beta=1,\ldots,s+2$.
We consider separately  electric and magnetic branes. In the first case,  $n=d$ and  the form field $F_{(n+1)} = dA_{(n)}$ is generated by the potential
\begin{equation}\label{eq:electric_ansatz}
    A_{01 \ldots d-1}=v(x). 
\end{equation}
In the magnetic case $n=s$, and the form field reads:
\begin{equation}\label{eq:magnetic_ansatz}
    F^{\alpha_1\ldots\alpha_{s+1}}
    =\frac{\e^{\alpha\phi}}{\sqrt{-G}} \epsilon^{\alpha_1\ldots\alpha_{s+1} \beta} \partial_\beta u(x).
\end{equation}
Here both the electric $v(x)$ and the  magnetic $u(x)$ potentials depend on the transverse coordinates as well.

It is convenient to rescale the world-volume metric as $g_{\mu\nu}=f^{2/d}\,\tilde{g}_{\mu\nu}$, where $f=\sqrt{-g}$, $\det{\tilde{g}_{\mu\nu}}=-1$.
In terms of new variables  the equations of motion reduce to equations of   the $s+2$-dimensional sigma model:
\begin{align}
S&=\frac{1}{2\kappa_D^2}\int\,d^{s+2}x\sqrt{h}\Big[R(h)-h^{\alpha\beta}
\Big(\frac{1}{2}\partial_\alpha\phi\partial_\beta\phi+
\frac{s+d}{sd}\partial_\alpha(\ln f)\partial_\beta(\ln f)\nonumber\\
\label{gen_sigma}
&+\frac14
\tilde g^{\mu\lambda}\partial_\alpha
\tilde g_{\lambda\nu}\tilde g^{\nu\sigma}\partial_\beta
\tilde g_{\sigma\mu}-\frac12{\rm e}^{-\psi}
\partial_\alpha v\partial_\beta v\Big)\Big].
\end{align}
where $R(h)$ is the Ricci scalar of the metric $h_{\alpha\beta}$ and $\psi=\alpha \phi+2 \ln f$.  The   matrix $\tilde g_{\mu\nu}$  decouples from the
rest of  variables, interacting with them only  through
the gravitational field $h_{\alpha\beta}$. Since $\tilde g_{\mu\nu}$ is
a symmetric matrix with (minus) unit determinant, this part of variables parametrizes a coset
$SL(d,R)/SO(1,d-1)$. Therefore the metric on the world-volume of
the $p$-brane is to high extent independent of the other $\sigma$-model
variables, which only influence its determinant. In the magnetic case one has to replace $v$ by $u$ and change the sign of the dilaton.

As result, we will get $s+2$-dimensional sigma-models on the transverse space with the metric $h_{\alpha\beta}$ realized by the target space variables  $f, \phi, \tilde{g}^{\mu\nu}$ and $v$ or $u$ respectively  for   electric ($+$) and magnetic ($-$) cases  \cite{Galtsov:1998mhf}.
The line elements of the corresponding target-spaces are
\begin{subequations}
\begin{equation} \label{eq:sigma_electric}
    dl_e^2 = A d\xi_{+}^2 + B d\psi_{+}^2 -\frac{1}{2} \e^{-\psi_{+}} dv^2
    + \frac{1}{4}\text{tr}\left[
        \tilde{g}^{-1} d\tilde{g}\,\tilde{g}^{-1} d\tilde{g}
    \right],
\end{equation}
\begin{equation} \label{eq:sigma_magnetic}
    dl_m^2 = A d\xi_{-}^2 + B d\psi_{-}^2 -\frac{1}{2} \e^{-\psi_{-}} du^2
    + \frac{1}{4}\text{tr}\left[
        \tilde{g}^{-1} d\tilde{g}\,\tilde{g}^{-1} d\tilde{g}
    \right],
\end{equation}
\end{subequations}
where $\xi_\pm, \psi_\pm$ are the following functions of $\phi$ and $f$
\begin{equation} \label{eq:xi_psi}
    \xi_{\pm} = \pm sd\phi - \alpha (s+d) \ln f, \qquad
    \psi_{\pm} = \pm \alpha \phi + 2\ln f,
\end{equation}
and $A, B$ are constants
\begin{equation*} \label{ABDelta}
    A = \frac{1}{sd\Delta},\qquad
    B = \frac{s+d}{2\Delta},\qquad
    \Delta = \alpha^2(s+d) + 2 sd.
\end{equation*}
There is an electromagnetic duality between the electric (\ref{eq:sigma_electric}) and magnetic (\ref{eq:sigma_magnetic}) sigma-models
\begin{equation} \label{eq:em_duality}
    \phi \to -\phi,\qquad
    u \leftrightarrow v,
\end{equation}
which translates an electric solution into a magnetic and vice versa.
Keeping this in mind, let us consider the electric case (\ref{eq:sigma_electric}).

The isometry group of the target space $SL(d,R)/SO(1,d-1) \times SL(2,R)/SO(1,1) \times R$ consists of the transformations of the matrix $\tilde{g}$, translations along $\xi_{+}$ and the non-trivial transformations in the subspace $(\psi_{+}, v)$ relevant for the generating technique.
Similarly to the four-dimensional Einstein-Maxwell model, one can define the Ernst potentials
\begin{equation}\label{eq:ernst}
    \Phi=\frac{v}{2\sqrt{2B}},\qquad
    \mathcal{E} = \exp{\psi_{+}}-\frac{v^2}{8B}.
\end{equation}
A non-trivial isometry transformation which preserves an asymptotic  behavior of the Ernst potentials $\mathcal{E} \to 1$, $\Phi \to 0$, is
\begin{equation}\label{eq:transformation}
    \Phi=\frac{
        \Phi^{(0)} + c( c \Phi^{(0)} + \mathcal{E}^{(0)} - 1)
    }{
        1 - 2c \Phi^{(0)} - c^2 \mathcal{E}^{(0)}
    },\qquad
    \mathcal{E} = \frac{
        \mathcal{E}^{(0)} + 2c \Phi^{(0)} - c^2
    }{
        1 - 2c \Phi^{(0)} - c^2 \mathcal{E}^{(0)}
    },
\end{equation}
where the index ${}^{(0)}$ stands for the seed solution   and $c$ is a real transformation parameter.
If the seed solution is uncharged $v_0=0$, the transformation (\ref{eq:transformation}) can be simplified as
\begin{equation}\label{eq:transformation_0}
    \Phi=c\frac{
        \mathcal{E}^{(0)} - 1
    }{
        1 - c^2 \mathcal{E}^{(0)}
    },\qquad
    \mathcal{E} = \frac{
        \mathcal{E}^{(0)} - c^2
    }{
        1 - c^2 \mathcal{E}^{(0)}
    }.
\end{equation}
The magnetic isometry transformations of the target space (\ref{eq:sigma_magnetic}) can be obtained by replacing $v$ and $\psi_{+}$ with $u$ and $\psi_{-}$ in (\ref{eq:ernst}). These transformations represent the generalized Harrison map for the model (\ref{eq:action}).

\section{Branes with independent scalar charge} \label{sec:charged_brane}
To generate a charged brane with an independent scalar charge, we will start with the seed solution of the model (\ref{eq:action}) with the trivial antisymmetric form. This is related to the FJNW solution in higher dimensions.
\subsection{FJNW solution in $D$-dimensional spacetime} \label{subsec:fisher}

Four-dimensional FJNW solution   was generalized to arbitrary dimensions by Xanthopoulos and Zannias (XZ) in Ref. \cite{Xanthopoulos:1989kb}. For the theory with the action
\begin{equation} \label{eq:action_fisher}
    \mathcal{S} = \frac{1}{16\pi}
    \int d^{\tilde{D}}x \sqrt{-g} \left(
        R - \frac{1}{2}(\nabla\phi)^2
    \right),
\end{equation}
the solution reads
\begin{align} \label{eq:fisher_solution}
    &
    ds^2 = -f_1^\sigma dt^2 + f_1^{\frac{1-\sigma}{\tilde{D}-3}-1}dr^2 + r^2 f_1^{\frac{1-\sigma}{\tilde{D}-3}}d\Omega^2_{(\tilde{D}-2)},\quad\quad
    \;\phi = \sqrt{\frac{\tilde{D}-2}{2(\tilde{D}-3)}}\frac{\Sigma \sigma}{M} \ln f_1,
    \\\nonumber &
    f_1 = 1 - \left(\frac{r_0}{r}\right)^{\tilde{D}-3},\quad
    r_0^{\tilde{D}-3} = \frac{ 16 M\pi }{ (\tilde{D}-2)\sigma S_{\tilde{D}-2} },\quad
    \sigma=\frac{M}{\sqrt{M^2+\Sigma^2}},
\end{align}
where the mass $M$ and the scalar charge $\Sigma$ are independent parameters, $S_n = {2\pi^{(n+1)/2}}/{\Gamma\left(\frac{n+1}{2}\right)}$ is the surface area of a unit $n$-sphere, $d\Omega^2_{(n)}$ is a line element of the unit $n$-sphere. One can directly check that the equations of motion for the model (\ref{eq:action_fisher}) correspond to the equations of motion (\ref{eq:eoms}) with vanishing antisymmetric form $F_{(n+1)} = 0$.
Thus, the FJNW solution satisfies the equations of motion for the model (\ref{eq:action}). The horizon at $r=r_0$ is always singular for $\Sigma\neq0$ \cite{Abdolrahimi:2009dc}. Zero scalar charge $\Sigma=0$ brings us back to the standard Schwarzschild-Tangherlini solution in $\tilde{D}$ dimensions.

\subsection{Generation of the charged FJNW-brane}
To get the charged FJNW-branes, one constructs the seed solution complementing the FJNW metric (\ref{eq:fisher_solution}) with a flat subspace parametrized by the coordinates $y_1,\ldots,y_{d-1}$.
The subspace metric in the sector $t, y_1, \ldots, y_{d-1}$ and the space-like section of the metric (\ref{eq:fisher_solution}) are identified with $g^{(0)}_{\mu\nu}$ and $h^{(0)}_{\alpha\beta}$ of the previous section respectively.
This gives us the seed potentials in the form $\psi_{+}^{(0)} = \sigma U \ln f_1$ and $\xi_{+}^{(0)} = \xi_{+} = \sigma V \ln f_1$, where
\begin{equation}
    U = \sqrt{\frac{s+1}{2s}} \frac{\alpha\Sigma_0}{M_0} + 1,\quad
    V = s d \sqrt{\frac{s+1}{2s}} \frac{\Sigma_0}{M_0} -\frac{1}{2}\alpha(s+d).
\end{equation}
Substituting $\psi_{+}^{(0)}$ and $\xi_{+}^{(0)}$ in (\ref{eq:ernst}), we will get the seed Ernst potentials ${\mathcal{E}^{(0)} = f_1^{\sigma U}}$, ${\Phi^{(0)}=0}$.
Then the transformation (\ref{eq:transformation_0}) leads to the new Ernst potentials
\begin{equation}
    \mathcal{E} = \frac{f_1^{\sigma U} - c^2}{1-c^2 f_1^{\sigma U}},\quad
    \Phi = c\frac{f_1^{\sigma U} - 1}{1-c^2f_1^{\sigma U}}.
\end{equation}
Using (\ref{eq:xi_psi}) and (\ref{eq:ernst}), one finds new   $v$, $\phi$ and $f$ for the generated solution. Finally, this leads  to the metric of the electric $p$-brane
\begin{align} \label{eq:charged_fisher_brane}
    &
    ds^2=
        f_2^{4s/\Delta} \left(
            -f_1^\sigma dt^2 + dy_1^2 + \ldots + dy_{d-1}^2 
        \right)
        +f_2^{-4d/\Delta} f_1^{(1-\sigma)/s} \left(
              dr^2/f_1 + r^2   d\Omega_{s+1}^2
        \right),
    \nonumber\\ &
    \phi = \frac{\sigma(U-1)}{\alpha} \ln f_1 + 4B\alpha \ln f_2,\qquad
    A_{01 \ldots d-1} =  2c\sqrt{2B}\left(f_1^{\sigma U}f_2-1\right),
\end{align}
where
\begin{align*}
    &
    f_1 = 1-\frac{r_0^s}{r^s},\qquad
    f_2 = \frac{1-c^2}{1 - c^2 f_1^{\sigma U}},
    \qquad
    r_0^s = \frac{16M_0\pi}{\sigma (s+1)S_{s+1}},\qquad
    \sigma = \frac{M_0}{\sqrt{M_0^2+\Sigma_0^2}}.
\end{align*}
This new solution is physically meaningful only for integers $s,\,d \geq 1$. The independent parameters are the seed mass   and scalar charge $ M_0,\,\Sigma_0$, as well as the Harrison transformation parameter $c$.   
The solution (\ref{eq:fisher_solution}) can be restored setting $c=0,\,d=1$.

The magnetic brane can be obtained via electric-magnetic duality (\ref{eq:em_duality}).
According to (\ref{eq:magnetic_ansatz}), the antisymmetric form will read
\begin{equation} \label{eq:magnetic_solution}
    F_{\alpha_1\ldots\alpha_{s+1}}
        = \frac{c}{1-c^2} \frac{s}{s+1} \frac{32 M_0 U \pi \sqrt{2B}}{ S_{s+1}}\sqrt{\Omega_{(s+1)}} \epsilon_{\alpha_1\ldots\alpha_{s+1}},
\end{equation}
where $\Omega_{(n)}$ is the determinant of the metric tensor of the unit $n$-sphere, $\epsilon_{\alpha_1\ldots\alpha_{s+1}} = \pm 1$.

Important examples of the generated family are F1 and NS5 branes in supergravities of type II. The fundamental string F1 is a solution with an electically charged 3-form with $d=2,s=6,\alpha=1$. Its dual one is a five-brane soliton NS5 with magnetic charge and $d=6,s=2,\alpha=1$.

\subsection{Charges}
The Komar mass $\mathcal{M}$ (defined with respect to the Killing vector $\mathcal{K}=\partial_t$) and the electric Page charge $\mathcal{Q}$ per unit area of the FJNW brane can be presented as the surface integrals
\begin{subequations}
\begin{equation} \label{eq:komar_mass1}
    \mathcal{M} = \frac{\kappa_D}{V_y} \int_{S_\infty\times M_p} \ast d\mathcal{K} = \frac{s}{s+1}M_0\left(1 + \delta_M\right),
\end{equation}
\begin{equation} \label{eq:komar_electric_charge1}
    \mathcal{Q} = \kappa_D \int_{S_\infty} \ast F^{(e)} = (-1)^d\frac{s}{s+1} \frac{c}{1-c^2} 2\sqrt{2B} M_0 U.
\end{equation}
\end{subequations}
where
\begin{equation} \label{eq:charges_constants1}
    \kappa_D = - \frac{1}{16\pi},\qquad
    \delta_M = \frac{4s}{\Delta}\frac{c^2}{1-c^2} U,
\end{equation}
where $V_y$ is a brane volume, $S_\infty$ is a hypersurface of an infinitely distant $s+1$-sphere, and $M_p$ is a spacelike section of the brane world-volume. In the magnetic case, the charge corresponding to the undualized form integral 
\begin{equation}
    \mathcal{P} = \kappa_D \int_{S_\infty} F^{(m)},
\end{equation}
formally coincides with the expression for the electric charge (\ref{eq:komar_electric_charge1}) up to sign.
 
In accordance with the dilaton asymptotics for  $r\to\infty$,
\begin{align} \label{eq:scalar_charge1}
    \phi \approx
    -\frac{1}{r^s}
    \frac{16\pi}{(s+1)S_{s+1}}
    \sqrt{\frac{s+1}{2s}}
    \Sigma_0
    \left( 1 + \delta_\Sigma \right),\qquad
    \delta_\Sigma =
    4B\alpha^2 \frac{U}{U-1}\cdot \frac{c^2}{1-c^2},
\end{align}
it is reasonable to define the dilaton charge $\mathcal{D}$ of the brane as follows
\begin{equation} \label{eq:new_scalar_charge}
    \mathcal{D} = - \sqrt{ \frac{s}{2 (s+1)} } \frac{ \Sigma_0 \left(1 + \delta_\Sigma\right) }{ \alpha } =
    - \left(
          1
        +
        \frac{4B\alpha^2c^2}{1-c^2}
    \right)\sqrt{ \frac{s}{2 (s+1)} } \frac{\Sigma_0}{\alpha}
    - \frac{4Bc^2}{1-c^2} \frac{s}{s+1} M_0 
    .
\end{equation}

In Sec. \ref{subsec:singularities} it will be shown that the necessary condition for the horizon to be regular is the seed scalar charge equal to zero: 
$\Sigma_0=0$.
This condition does not mean a zero dilaton charge $\mathcal{D}=0$, since in this case $U=1$, and the product $\Sigma_0\, \delta_\Sigma$ is finite. Rather, in this limit we obtain a constraint
\begin{equation} \label{eq:regularity_condition_electric}
    \mathcal{Q}^2 = \mathcal{D} \left(\frac{ \Delta - 4s }{d+s} \mathcal{D} - 2 \mathcal{M}\right)
\end{equation}
for the electric brane and
\begin{equation} \label{eq:regularity_condition_magnetic}
    \mathcal{P}^2 = \mathcal{D} \left(\frac{ \Delta - 4s }{d+s} \mathcal{D} + 2 \mathcal{M}\right)
\end{equation}
for the magnetic one (which can be obtained replacing $\mathcal{D}\to-\mathcal{D}$, $\mathcal{Q} \to \mathcal{P}$).
For the parameter values $d=s=1$, $\alpha=\sqrt{3}, (\Delta - 4s)/(d+s) = 2$, the model is the four-dimensional $\alpha=\sqrt{3}$ Einstein-Maxwell-dilaton (EMD) theory, and the conditions (\ref{eq:regularity_condition_electric}), (\ref{eq:regularity_condition_magnetic})  coincide with the charge constraint found by Rasheed \cite{Rasheed:1995zv}. 
The Eq. (\ref{eq:regularity_condition_electric}) has two roots with respect to $\mathcal{D}$:
\begin{equation} \label{eq:regularity_resolved}
    \mathcal{D}_{\pm} = \frac{ d + s }{ \Delta - 4s } \mathcal{M} \left(1 \pm \sqrt{1 + \frac{ \Delta - 4s }{d+s} \cdot \frac{\mathcal{Q}^2}{\mathcal{M}^2} }\right).
\end{equation}
If $\mathcal{Q}=0$, the regularity condition (\ref{eq:regularity_condition_electric}) should coincide with the regularity condition of the seed solution, namely $\Sigma_0 = \mathcal{D} = 0$.
Therefore, the root $\mathcal{D}_{+}$ is unphysical and does not correspond to the constraint $\Sigma_0=0$.
Similarly, the plus sign in the solution for Eq. (\ref{eq:regularity_condition_magnetic}) does not correspond to regular magnetic solutions.

Another interesting limit is $U\to0$, $c^2\to 1$. In this case the expressions for charges (\ref{eq:komar_mass1}), (\ref{eq:komar_electric_charge1}), (\ref{eq:new_scalar_charge}) and the solution itself (\ref{eq:charged_fisher_brane}) have a non-trivial limiting form with new metric function
\begin{equation}\label{eq:limit_1}
    f_2 = \left(\zeta \ln f_1 + 1\right)^{-1},
\end{equation}
where $\zeta$ is a new independent parameter.
For $\zeta > 0$, the function $f_2$ diverges at $r_\zeta=r_0\left(1-e^{-1/\zeta}\right)^{-1/s} > r_0$ as $1/(r-r_\zeta)$.
For $\zeta < 0$, the function $f_2$ does not have roots  and tends to zero at $r_0$ as $1/\ln f_1$. 

Similarlly, the limit $M_0,\,\Sigma_0 \to 0,\,c^2 \to 1$ leaves the charges finite.
Let us introduce an infinitesimal parameter $\epsilon$, such that the quantities $M_0,\,\Sigma_0,\,c$ depend on it as $ M_0 = \xi \epsilon,\, \Sigma_0 = \zeta \epsilon,\, c = \pm \left(1 - \epsilon\right)$, where $\zeta$ and $\xi$ are some constants.
Then, the limit $\epsilon \to 0$ leads to   redefinition
\begin{equation}\label{eq:limit_2}
    f_1 = 1,\qquad
    f_2 = \frac{r^s}{\rho^s + r^s},
\end{equation} 
and the new charges will be
\begin{equation}\label{eq:limit_E_charges}
    \mathcal{M} = \frac{2 s \Phi}{\Delta},\quad
    \mathcal{D} = -\frac{\Phi(d+s)}{\Delta},\quad
    \mathcal{Q} = \pm\Phi\sqrt{\frac{d+s}{\Delta}},
\end{equation}
where
\begin{equation*}
    \rho^s = \frac{ 8 \pi \Phi }{ s S_{s+1} },\qquad
    \Phi = 
    \frac{s\xi}{s+1} + \alpha  \zeta \sqrt{\frac{s}{2(s+1)}}.
\end{equation*}
In the magnetic case, the solution can be found from (\ref{eq:limit_E_charges}) again replacing $\mathcal{D}\to-\mathcal{D},\,\mathcal{Q}\to \mathcal{P}$.
It can be verified that the charge expressions satisfy identically the  condition (\ref{eq:regularity_resolved}).  
These solutions were found previously in Ref. \cite{Lu:1995cs}. The charges (\ref{eq:limit_E_charges}) satisfy the following constraints
\begin{subequations}
    \begin{equation}\label{eq:mass_constraint}
        \mathcal{M}^2 = \mathcal{Q}^2 \frac{4s^2}{(d+s)\Delta},
    \end{equation}
    \begin{equation}\label{eq:scalar_constraint}
        \mathcal{D}^2 = \mathcal{Q}^2 \frac{d+s}{\Delta},
    \end{equation}    
\end{subequations}
which can be combined into
\begin{equation}\label{eq:no_force_like}
    \mathcal{M}^2 + \left( \alpha^2 + 2s\frac{d(d+s) - 2s}{(d+s)^2} \right) \mathcal{D}^2 - \mathcal{Q}^2 = 0,
\end{equation}
resembling the no-force condition  \cite{Scherk:1979aj,Tseytlin:1996hi}.
For $d=s=1$ the mass constraint (\ref{eq:mass_constraint}) coincides with the known result $\mathcal{M}^2 = \mathcal{Q}^2/(1+\alpha^2)$ for BPS-solutions in the EMD model \cite{Rasheed:1995zv, Nozawa:2010rf,Poletti:1995yq,Geng:2018jck}.
    
The limit $c\to\infty$ leads to the function redifinition
\begin{equation}\label{eq:limit_3}
f_2 = f_1^{-\sigma U},
\end{equation}
while the antisymmetric form becomes trivial.
This solution is a particular case of the wider family, which can be obtained from the seed solution with $SO(2)$ transformation considered in Sec. \ref{sec:zv}.

The first limiting solution (\ref{eq:limit_1}) does not differ qualitatively from the general family, while the second limiting solution (\ref{eq:limit_3}) is a special case of another wider family of uncharged solutions that requires a separate study. 
In what follows, we will discuss physical property of the  general solution (\ref{eq:charged_fisher_brane})  and the second limiting solution (\ref{eq:limit_2}),  denoting them as $S_G$ and $S_E$ respectively.

\subsection{Singularities } \label{subsec:singularities}
To perform an analysis of spacetime structure, we  calculate the Ricci scalar, contracting indices in Eq. (\ref{eq:eom_gravity}), and substituting the general solution for electric/magnetic forms and the scalar field into the energy-momentum tensor
\begin{align} \label{eq:ricci}
    & R = 
    \frac{1}{2}\sigma^2
    f_2^{4d/\Delta} f_1^{(\sigma-s-1)/s}
    {f'_1}^2
    \left( R_F + R_\phi \right),\\\nonumber
    & R_F = \frac{
            4c^2 U^2 (d-s) f_2^2 f_1^{\sigma U}
        }{
            (1-c^2)^2\Delta
        },\\\nonumber
    & R_\phi = \left(
        \frac{\Sigma_0}{M_0}\sqrt{\frac{s+1}{2s}}
        + \frac{2\alpha c^2 U(d+s) f_2 f_1^{\sigma U}}{(1-c^2)\Delta}
    \right)^2.
\end{align}
For the general family $S_G$, the function $f_2$ is regular at $r=r_0$, therefore this factor does not lead to the curvature singularity.
But for $\Sigma_0\neq0$, the Ricci scalar (\ref{eq:ricci}) contains a diverging factor proportional to $f_1^{(\sigma - s - 1)/s}$ due to the negative exponent.
For $\Sigma_0 = 0$, $M_0>0$,   the surface $r=r_0$ is regular   and represent an event horizon if $|c| < 1$.

The function $f_2$  diverges once its denominator is zero:   $f_1^{\sigma U} = c^{-2}$.
The function $f_1$ is bounded from above by a value $1$, thus this is possible for $c^2>1$ only.
From Eq. (\ref{eq:ricci}) follows that this leads to the existence of one more singularity at $r_0(1-c^{-2/\sigma U})^{-s}>r_0$.

In the limiting case $S_E$, the scalar curvature becomes
\begin{equation} \label{eq:limit3_ricci}
    R = 
    \left(
        \frac{ 2 s \rho^s }{ r^{s+1}\Delta }
    \right)^2
    f_2^{4d/\Delta+2} 
    d (\Delta - s (d+s))
    \sim
    r^{4ds/\Delta-2}
    .
\end{equation}
To find out whether the point $r=0$ is singular, we  look at the extrema of the exponent $-2 + 4ds/\Delta$ with respect to $\alpha^2$.
This expression takes the lowest value $-2$ for $\alpha^2\to\infty$, and the largest value $0$ for $\alpha^2 = 0$.
So, the family $S_E$ is singular for $\alpha\neq 0$.
For $\Delta - s (d+s)=0$ (i.e. $\alpha^2 = s(s-d)/(s+d)$, $s \geq d$) the metric tensor is Ricci-flat.
However, the direct calculation of the scalars $R_{MN}R^{MN}$ and $R_{MNLK}R^{MNLK}$ lead to the asymptotic $\sim r^{4(d-s)/(d+s)}$ as $r \to 0$, so they diverge if $s \neq d$ (the case $s=d$ brings us back to $\alpha=0$).

The outermost singularity/horizon surface for $S_G$, $c^2 < 1$ is $r=r_0$.
Solutions with positive (negative) seed mass $M_0$ have an infinite red shift $g_{tt}\to 0$ (blue shift $g_{tt}\to\infty$)  at $r=r_0$. In Sec. \ref{sec:geodesics}, we will show that the the point $r=r_0$ is approached by the radial geodesic  in a finite time  of the remote observer for any $\sigma\neq1$.
The outermost interesting surface for $S_G$, $c^2>1$ is the surface $f_2^{-1} = 0$.
As $g_{tt}\to\infty$, such solutions are naked singularities with an infinite blue shift.
The family $S_G$ cannot be extremal in the sense $g'_{tt}(r_0)\neq0$ or ${g^{rr}}'(r_0)\neq 0$, which can be verified directly.

According to this analysis, the generic family $S_G$ is a naked singularity for $\sigma\neq1$, but  may have some black hole properties (redshift). The class $S_E$ represent extreme solutions, that are regular only for $\alpha=0$, representing extreme Reissner-Nordstr\"om singly charged solutions.

\subsection{Supersymmetry}
The fact that singularities of spacetime may be compatible with supersymmetry is known from the example of domain walls 
\cite{Kallosh:2001zd}. Therefore, it makes sense to check the possible supersymmetry of our singular branes from first principles. 
It can be seen that the solution subfamily $S_E$  contains the known family of the BPS saturating solutions for any $d$, $s$, $\alpha$, representing $p$-branes of the type IIA/IIB superstring theories \cite{Dabholkar:1990yf,Duff:1994an,Clement:2004ii}. However, the question may arise whether the general solution $S_G$ contains other supersymmetric solutions. 

We will study supersymmetry in the case of NS5-brane ($d=6$, $s=2$, $\alpha=1$) in type IIA supergravity. The supersymmetry conditions have the simplest form in the string frame $G_{MN}^{(s)} = e^{-\phi/2}G_{MN}$:
\begin{align}\label{eq:solution_ns5}
    &
    ds^2_{(s)}=
        f_1^{-\sigma(U-1)/2} \left(
            -f_1^\sigma dt^2 + dy_1^2 + \ldots + dy_{5}^2 
        \right)
        +f_2^{-1}f_1^{-(1+\sigma U)/2} \left(
            dr^2 + r^2 f_1 d\Omega_{3}^2
        \right),
    \\\nonumber &
    \phi = -\sigma(U-1) \ln f_1 - \frac{1}{2} \ln f_2.
\end{align}
We will choose the three-sphere metric as follows
\begin{equation}
    d\Omega_{3}^2 = d\psi^2 + \sin^2\psi \left(
        d\theta^2 + \sin^2\theta d\varphi^2
    \right),\qquad
    \sqrt{\Omega_{(3)}} = \sin^2\psi \sin\theta,
\end{equation}
so the magnetic three-form from the Neveu-Schwarz sector $H_{MNP}$ reads
\begin{equation}
    H_{\psi\theta\varphi} = p \sin^2\psi \sin\theta,\qquad
    p = \frac{c}{1-c^2} \frac{16 M_0 U}{ 3 \pi},
\end{equation}
where we introduced the constant $p$ for convenience.

The variations of gravitino and dilatino   for type IIA supergravity with truncated Ramond-Ramond fields are \cite{Becker:2007zj, Kallosh:2001ag}
\begin{subequations}
    \begin{equation} \label{eq:gravitino_general}
        \delta\psi_M = \left(\nabla_M - \frac{1}{8} H_{MNP} \Gamma^{NP} \Gamma_{11} \right)\epsilon,
    \end{equation}
    \begin{equation} \label{eq:dilatino_general}
        \delta\lambda = -\frac{1}{3} \left(
              \Gamma^\mu \left(\partial_\mu \phi\right) \Gamma_{11}
            - \frac{1}{12} H_{MNP} \Gamma^{MNP}
        \right)\epsilon,
    \end{equation}
\end{subequations}
where
\begin{equation}
    \nabla_\mu = \partial_\mu + \frac{1}{4} \omega_{MAB}\Gamma^{AB},\qquad
    \Gamma^{M_1 \ldots M_n} = \Gamma^{[M_1}\ldots\Gamma^{M_n]},\qquad
    \Gamma_{11} = \Gamma_0 \Gamma_1 \ldots \Gamma_9,
\end{equation}
$\omega_{MAB}$ is spin connection, and we use the convention $\{\Gamma_M, \Gamma_N\} = 2 G^{(s)}_{MN}$.

Since the Killing spinor does not depend on the coordinates along the orbits of the commuting Killing vectors $y^\mu$, so the condition $\delta \psi_\mu = 0$ reduces to ${\omega_{\mu}}^{\bar \nu \bar r} \Gamma_{\bar \nu \bar r} \epsilon = 0$, where the flat indices with respect to the ten-dimensional metric $G_{MN}^{(s)}$ are denoted by bars. As gamma matrices are not degenerate, this requires the spin connection to be zero, which is possible if $G^{(s)}_{\mu\nu}$ is constant (for $\mu,\,\nu=0,\ldots 5$ only). This is not possible for the general solution $S_G$ unless the function $f_1$ is constant. The function $f_1$ is constant for the subfamily $S_E$ only with $f_1 = 1$ and $f_2 = r^s/(r^s + \rho^s)$, and there is no other solutions in $S_G$ satisfying this condition. As we have mentioned, this solution is known to have Killing spinors. For the completeness of the analysis we will obtain the resulting Killing spinors. The component $\delta \psi_r = \partial_r \epsilon$ is zero if the Killing spinor $\epsilon$ depends on the angles $\psi$, $\theta$ and $\varphi$ only. The remaining gravitino equations on the three-sphere are solved by the spinor
\begin{equation}
    \epsilon = 
    \text{exp}\left(-\Gamma^{\bar\psi}\Gamma^{\bar r}\psi/2\right)
    \text{exp}\left(-\Gamma^{\bar\theta}\Gamma^{\bar \psi}\theta/2\right)
    \text{exp}\left(-\Gamma^{\bar\varphi}\Gamma^{\bar \theta}\varphi/2\right)
    \epsilon_0,
\end{equation}
where $\epsilon_0$ is a constant spinor and the exponents can be expanded as
\begin{equation}
    e^{\Gamma^M \Gamma^N x } = \cos x + \Gamma^M \Gamma^N \sin x,\qquad
    M \neq N \neq t.
\end{equation}
The remaining dilatino equation (\ref{eq:dilatino_general}) for $S_E$ solutions is
\begin{equation}
    (1 \pm \Gamma_0 \Gamma_1 \Gamma_2 \Gamma_3 \Gamma_4 \Gamma_5) \epsilon
    = 0,
\end{equation}
where the sign depends on the sign of the magnetic charge. This condition breaks a half of supersymmetry.

\subsection{FJNW-branes with oblateness parameter}\label{sec:zv}
Static uncharged solutions in General Relativity with minimally coupled scalar field, which is the truncation of the theory (\ref{eq:action}) with $F=0$, can be reduced to a simple sigma-model.
This opens a way to add the minimal massless scalar field via the transformations.  

To construct an appropriate sigma-model, we can truncate the previously derived  ones (\ref{eq:sigma_electric}) or (\ref{eq:sigma_magnetic}) setting zero electric $v$ or magnetic $u$ potentials. This gives the target space metric
\begin{equation} \label{eq:sigma_zv}
    dl_0^2 = \frac{s+d}{sd}\left(d \ln \sqrt{-g}\right)^2 + \frac{1}{2} d\phi^2.
\end{equation}
It   has an $SO(2)$ symmetry, corresponding to the transformation used in \cite{Abdolrahimi:2009dc}
\begin{equation}\label{eq:so2_transformations}\begin{cases}
 \ln\sqrt{-g} = 
 \cos\beta \cdot \ln \sqrt{-g^{(0)}}
 - \sqrt{\frac{sd}{2(s+d)}}\sin\beta \cdot \phi^{(0)}\\
 \phi = 
 \sqrt{\frac{2(s+d)}{sd}}\sin\beta \cdot \ln \sqrt{-g^{(0)}} 
 + \cos\beta \cdot \phi^{(0)}
\end{cases},\end{equation}
where $\beta$ is a transformation parameter.

Starting with the four-dimensional vacuum axially symmetric Zipoy-Voorhees solution \cite{Zipoy:1966,Voorhees:1971wh}, the $p$-branes with $s=1$ can be supplied with the deformation (oblateness) parameter $\delta$.
One chooses the seed solution 
\begin{align} \label{eq:ZV_metric}
    & ds^2 =-f_1(x)^\delta dt^2 + f_1(x)^{-\delta} ds_{(3)}^2, \qquad
    f_1(x) = \frac{x-1}{x+1},
    \\\nonumber
    & ds_{(3)}^2 = k^2 \left[
        \left(
            \frac{x^2-1}{x^2-y^2}
        \right)^{\delta^2-1}
        \left(
            dx^2 + \frac{x^2-1}{1-y^2} dy^2
        \right)
        +(x^2-1)(1-y^2)d\varphi^2
    \right],
\end{align}
where $k$ and $\delta$ are some constants, $x$ and $y$ are oblate spheroidal coordinates, related to spherical coordinates as $x = \frac{r}{k} - 1, \, y = \cos\theta$.
Substituting $d=s=1$, $|g^{(0)}|=f^\delta$ and $\phi^{(0)}=0$ in $SO(2)$-transformations (\ref{eq:so2_transformations}) and redefining $\beta$ through the charges, we will get a new solution
\begin{equation} \label{eq:fzv_metric}
    ds^2= -f_1^{\delta\sigma} dt^2 + f_1^{-\delta\sigma} ds_{(3)}^2,\qquad
    \phi = \frac{\Sigma}{k} \ln f_1,
\end{equation}
where $\sigma$ has the former expression, $k=\cfrac{M}{\sigma\delta}$.
The obtained solution combines FJNW and Zipoy-Voorhees four-dimensional solutions.
Next, we can extend the solution to $D=d+3$ dimensions and apply the transformations (\ref{eq:transformation_0}).
We will get   solution in the form of the  previously found FJNW brane   (\ref{eq:charged_fisher_brane}) with the replacement $f_1^{\sigma} \to f_1^{\sigma\delta}$ and using the new form of $ds_{(3)}$: 
\begin{align} \label{eq:charged_FZV_brane}
    &
    ds^2=
        f_2^{4/\Delta} \left(
            -f_1^{\sigma\delta} dt^2 + dy_1^2 + \ldots + dy_{d-1}^2 
        \right)
        +f_2^{-4d/\Delta} f_1^{-\sigma\delta} ds^2_{(3)},
    \\\nonumber &
    \phi(r) = \frac{\Sigma}{k} \ln f_1
            + 4B\alpha \ln f_2,\qquad
    f_1(r) = \frac{x-1}{x+1},\qquad
    f_2(r) = \frac{1-c^2}{1 - c^2 f_1^{\sigma\delta U}},
    \\\nonumber &
    \sigma = \frac{M}{\sqrt{M^2+\Sigma^2}},\qquad
    U = 1 + \frac{\Sigma \alpha}{M},
\end{align}
with the electric potential
\begin{equation*}
    A_{01 \ldots d-1} = \frac{2\sqrt{2B}}{c} \left(f_2-1\right)
\end{equation*}
or the magnetic form
\begin{equation*}
    F_{y\varphi} = \frac{4c\sqrt{2B}}{1-c^2}\left(M+\alpha\Sigma\right).
\end{equation*}
This is a solution with both an independent scalar charge and an arbitrary oblateness parameter $\delta$  (see Ref. \cite{Bogush:2020lkp} for more details of the replacement $\sigma\to\sigma\delta$).

\section{Geodesics}\label{sec:geodesics}

Consider the geodesic equation
\begin{equation} \label{eq:geodesics_eq}
    \frac{d}{d\tau} \left(G_{MN} \dot{X}^N\right)
    - \frac{1}{2} \frac{d G_{PQ}}{dX^M} \dot{X}^P \dot{X}^Q = 0.
\end{equation}
The metric (\ref{eq:charged_fisher_brane}) can be written in the form
\begin{align} \label{eq:geodesics_metric}
    ds^2 &= - a(r) dt^2
        + b(r) \left.\bm{dy}\right.^2
        + v(r) dr^2 
        + w(r) d\Omega_{s+1}^2,
\end{align}
where
\begin{align} \label{eq:geodesics_metric_params}
    & a(r) = f_2^{4s/\Delta}f_1^\sigma,\quad
    & b(r) = f_2^{4s/\Delta},
    \\\nonumber
    & v(r) = f_2^{-4d/\Delta} f_1^{(1-s-\sigma)/s},\quad
    & w(r) = f_2^{-4d/\Delta} f_1^{(1-\sigma)/s} r^2.
\end{align}
By virtue of the spherical symmetry, we can choose geodesics lying in the equatorial plane of $s+1$-sphere along the coordinate $\varphi$.
Constants of motion for $t$, $y^i$ and $\varphi$ follows from Eq. (\ref{eq:geodesics_eq})
\begin{equation} \label{eq:geodesics_integrals}
    \dot{t} = E a^{-1}, \qquad
    \dot{y}^i = k^i b^{-1}, \qquad
    \dot{\varphi} = L w^{-1}.
\end{equation}
Let's choose the affine parameter $\tau$ so that $\dot{X}_M \dot{X}^M = - \epsilon$, where $\epsilon = -1,\,0,\,1$  for spacelike, lightlike and timelike geodesics respectively. Keeping in mind the constants of motion (\ref{eq:geodesics_integrals}), the radial equation reads
\begin{equation}  \label{eq:geodesics_radial_tau}
    \dot{r}^2 =
    \frac{1}{av}\left(
        E^2
        - V_{\text{eff}}
    \right),
\end{equation}
\begin{equation}\label{eq:geodesics_veff}
    V_{\text{eff}}(r) =
        \frac{a}{b}{\bm k}^2 + \frac{a}{w}L^2 + a\epsilon,
\end{equation}
where ${\bm k}^2$ is $k^i k^j \delta_{ij}$. In terms of the time of an infinitely distant observer, the radial equation has the form
\begin{equation}\label{eq:geodesics_radial_t}
    \left(\frac{dr}{dt}\right)^2 =
    \frac{a}{E^2 v}\left(
        E^2 - V_{\text{eff}}
    \right).
\end{equation}

The proper time  and the  distant observer time   of motion from $r_1$ to $r_2$, are
\begin{equation} \label{eq:proper_observer_time_near_horizon}
    \Delta\tau = \pm\int_{r_1}^{r_2} \frac{\sqrt{av} dr}{\sqrt{E^2 - V_\text{eff}}},\qquad
    \Delta t = \pm\int_{r_1}^{r_2} \sqrt{\frac{v}{a}}\frac{dr}{\sqrt{1 - V_\text{eff}/E^2}}
\end{equation}

\begin{table}[ht]  \caption{behavior of $a$, $a/b$, $a/w$ and $\sqrt{av}$, $\sqrt{v/a}$ near the surface $S$ of the radius $r_S$ up to a multiplicative constant, $x = r - r_S$.\label{tab:geodesics_1}}
\begin{center}\begin{tabular}{ |c|c|c|c|c| }
 \hline
 Quantity & General expression & $S_G,\,c^2<1$ & $S_G,\,c^2>1$ & $S_E$ \\
 \hline\hline
 $r_S$ & & $r_0$ & $r_0(1-c^{-2/\sigma U})^{-s}$ & $0$ \\
 $a$ & $f_2^{4s/\Delta} f_1^{\sigma}$ & $x^{\sigma}$ & $x^{-4s/\Delta}$ & $x^{4s^2/\Delta}$ \\
 $a/b$ & $f_1^\sigma$ & $x^\sigma$ & $x^0$ & $1$ \\
 $a/w$ & $f_2^{4(s+d)/\Delta}f_1^{\sigma - \frac{1-\sigma}{s}}r^{-2}$ & $x^{\sigma - \frac{1-\sigma}{s}}$ & $x^{-4(s+d)/\Delta}$ & $x^{-2+4s(s+d)/\Delta}$ \\
 $\sqrt{av}$ & $f_2^{2(s-d)/\Delta} f_1^{\frac{(1-s)(1-\sigma)}{2s}}$ & $ x^{\frac{(1-s)(1-\sigma)}{2s}}$ & $x^{2(d-s)/\Delta}$ & $x^{2s(s-d)/\Delta}$ \\
 $\sqrt{v/a}$ & $f_2^{-2(s+d)/\Delta} f_1^{\frac{-(s+1)\sigma + 1-s}{2s}}$ & $x^{\frac{-(s+1)\sigma + 1-s}{2s}}$ & $x^{2(s+d)/\Delta}$ & $x^{-2s(s+d)/\Delta} $\\
 \hline
\end{tabular}\end{center}
\end{table}
Geodesics can reach the spherical surface $S$ with radius $r_S$ if the effective potential $V_{\text{eff}}$ does not tend to $+\infty$ at this surface.
For null and time-like worldlines, the effective potential is a positive function in the exterior region of the solution.
Hence, it is enough to require $V_\text{eff}$ to be bounded.
If every term of the effective potential behaves as $\sim(r-r_S)^a$ near $S$, where $a$ is some constant, then the necessary and sufficient condition is $a \geq 0$ for every term in the effective potential.
If the geodesic curve intersects the surface $S$, then the required time interval in some frame of reference can be finite or infinite. 
For a distant observer, the geodesic reaches the surface $S$ in a finite time interval $\Delta t$ if and only if the quanity $\sqrt{v/a}$ is integrable near this surface.
Therefore, if we want the time interval to be finite, the function $\sqrt{v/a}$ cannot diverge as $(r-r_S)^{-1}$ or faster.
Similarly, for the proper time $\Delta \tau$, we have a condition that the function $\sqrt{va}$ cannot diverge as $(r-r_S)^{-1}$ or faster.
Table \ref{tab:conditions} provides the corresponding condition for the functions $a$, $a/b$, $a/w$, $\sqrt{va}$ and $\sqrt{v /a}$, based on the results of Table \ref{tab:geodesics_1}.

\begin{table}[ht] \caption{Regularity condition of various terms in effective potential (\ref{eq:geodesics_veff}) and conditions of finiteness of the required time for geodesics to traverse the surface $S$ of the   radius $r_S$ in terms of the proper time $\tau$ and the distant observer time   $t$ for different solutions; $e_\epsilon$, $e_k$, $e_L$, $e_\tau$, $e_t$ are the exponents of the leading term in the expansion of the functions $a$, $a/b$, $a/w$, $\sqrt{av}$, $\sqrt{v/a}$ near the surface $S$ respectively. \label{tab:conditions}}
\begin{center}\begin{tabular}{ |c||c|c|c|c|c|c| }
 \hline
 Solution &
 $r_S$ &
 $e_\epsilon \geq 0$ &
 $e_k \geq 0$ &
 $e_L \geq 0$ &
 $e_\tau > -1$ &
 $e_t > -1$
 \\
 \hline\hline
 $S_G,\,c^2<1$ & $r_0$ & $\sigma\geq 0$ & $\sigma\geq 0$ & $\sigma\geq\frac{1}{1+s}$ & Always & $\sigma \neq 1$ \\
 $S_G,\,c^2>1$ & $r_0(1-c^{-2/\sigma U})^{-s}$ & Never & Always & Never & Always & Always \\
 $S_E$ & 0 & Always & Always & $\alpha^2 \leq \frac{2s^2}{s+d}$ & Always & $\alpha^2 > \frac{2s^2}{s+d}$\\
 \hline
\end{tabular}\end{center}
\end{table}

Behavior of geodesics in the background of solutions $S_G$ with $c^2<1$ does not drastically differs from the motion in uncharged FJNW-branes.
Depending on the value of $\sigma$, the effective potential can be either bounded or divergent in the singularity.
For $\sigma \geq 1/(1+s)$ the effective potential is always bounded for any geodesics; for $0 \leq \sigma < 1/(1+s)$ the potential is bounded for geodesics with zero angular momentum   $L=0$ only; in the case $\sigma < 0$, the singularity $r_0$ is reachable for null radial geodesics only ($\epsilon=k=L=0$).
Geodesics always traverse the surface $r_0$ in a finite proper time $\Delta\tau$.
The distant observer can observe this in a finite time $\Delta t$ only in the singular case $\sigma\neq1$.

Solutions $S_G$ with $c^2>1$ have another outermost singularity, which has different  properties for geodesics.
Timelike geodesics never reach this singularity due to an unbounded growth of the effective potential.
Null geodesics can reach the singularity only if they are purely radial, $L=0$.
From the point of view of the distant observer, null geodesics can reach the singularity in a finite time $\Delta t$ and  a finite   $\Delta \tau$.

The effective potential of geodesics in the background of $S_E$ with $\alpha^2 \leq \alpha^2_\text{crit}$, $\alpha^2_\text{crit} = 2s^2/(s+d)$ is bounded.
Geodesics traverse the surface $r=0$ in a finite $\Delta\tau$, but an infinite $\Delta t$.
For the other case $\alpha^2 > \alpha^2_\text{crit}$, the effective potential with a nonzero angular momentum  $L$  unboundedly grows at the singularity, and geodesics with zero angular momentum $L=0$ reaches the singularity with finite   $\Delta t$ and $\Delta \tau$. In the case of the NS5-brane, the dilaton coupling constant has a critical value $\alpha = \alpha_\text{crit} = 1$. 

\section{Test scalar field}
 
The physical effects of the field theory that can manifest themselves in a singular spacetime are interesting both from the point of view of the search for new physics \cite{Liao:2014vpa,Chowdhury:2020rfj} and from a theoretical point of view.  First of all one has to construct
the self-adjoint wave operators which is  a non-trivial task in singular curved spacetimes, since the usual axioms of the quantum field theory break there \cite{Wald:1980jn}.
This is an important question, since it is believed that quantum effects are of decisive importance near singularities.

A related issue is the possible quantum unobservability of classical singularities. It was argued that if there is a unique self-adjoint extension of the wave operator in a singular spacetime, then classical singularity is quantum-mechanically unobservable  by the corresponding field \cite{Wald:1980jn, Horowitz:1995gi}. This issue was investigated in a number of subsequent works.
The situation simplifies in the static spaces.   There are two simple ways to check the essential self-adjointness of symmetric operators. The first is the calculation of the defect indices (von Neumann). The second one uses Weyl's theorem: the wave operator is not self-adjoint if both local solutions in a neighborhood of the singular submanifold satisfy the criterion of square integrability ({\em limit circle}). Indeed, in this case, it is necessary to make additional assumptions about boundary conditions on which the evolution of the solution will depend. If only one solution is locally square-integrable ({\em limit point}), the evolution of the waves will be uniquely determined, so the wave packet will not feel the presence of the singularity. In this case, in the vicinity of the singularity, as a rule, a repulsive barrier arises. The idea was formulated by Wald \cite{Wald:1980jn} and more specifically implemented by Horowitz and Marolph \cite{Horowitz:1995gi} in the case of the scalar field. Later work in this direction includes Refs. \cite{Ishibashi:1999vw,Helliwell:2003tx,Ishibashi:2003jd} and others. The scalar theory was generalized to the Maxwell and Dirac fields  \cite{Helliwell:2003tx}.
The possibility of quantum healing of geodesically incomplete manifolds has been tested for a number of static and conformally-static singular solutions of Einstein theory in   four  and three dimensions (see the recent review \cite{Konkowski:2018imc}).

A closely related problem is that of stability of singular spacetimes. 
Stability of a solution is related to the sign of the imaginary part of the frequencies of normal modes defined with appropriate boundary conditions in singularities.   It may happen that different choice of boundary conditions in the case where both local solutions are square-integrable, may lead to different conclusions about stability~\cite{Gibbons:2004au,Sadhu:2012ur}. Physically this means that the full specification of the singular spacetime must include prescription of boundary conditions for perturbations in singularities. This issue requires further study which is beyond the scope of the present work. But the crucial question is the number of local solutions near the singularities which are square integrable.

In this section we  investigate possibility of quantum healing of singularities using  the test scalar field of mass $\mu$ on the background of new singular branes. Using the definitions (\ref{eq:geodesics_metric_params}), the corresponding Klein-Gordon equation $(\Box-\mu^2)\phi=0$ can be written in the form
\begin{equation} \label{eq:dalambertian}
    -a^{-1} \partial^2_t \phi
    +
    b^{-1} \partial^2_i \phi
    +
    \frac{1}{R} \partial_r \left(
        R v^{-1} \partial_r \phi
    \right)
    +
    \frac{1}{w}\Delta_{S}^{s+1} \phi
    -\mu^2\phi= 0,
\end{equation}
where $\Delta_{S}^{s+1}$ is the Laplace-Beltrami operator of a unit $s+1$-sphere, and
\begin{equation*}
    R(r) = \sqrt{-G/\Omega_{(s+1)}} = 
    \sqrt{a b^{d-1} v w^{s+1}} = f_2^{-4d/\Delta}f_1^{\frac{1-\sigma}{s}}r^{s+1}.
\end{equation*}
Substituting the ansatz $\phi = \exp\left\{-i\omega t + i k_i y^i\right\} \Phi(r) Y_l^K(\theta)$, where $K$ is a set of quantum numbers of the $s+1$-sphere harmonics $Y^K_l$, except for the quantum number of the total angular moment $l$. Taking into account the eigenvalues of the harmonics $\Delta_{S}^{s+1} Y_l^K = - l (l + s) Y_l^K$ (the eigenvalue $l$ is degenerate with the degree $(2l + s) (l+s-1)!/l!s!$  \cite{Harnad:1994wd}), we will get
\begin{equation} \label{eq:dalambertian3}
    \frac{1}{R} \partial_r \left(
        R v^{-1} \partial_r \Phi
    \right)
    + \left(
        \frac{\omega^2}{a}
        - \frac{{\bm k}^2}{b}
        - \frac{l(l+s)}{w}
        - \mu^2
    \right)\Phi = 0.
\end{equation}
Let us make a coordinate transformation $\partial_r = W(\varrho) \partial_\varrho$ and substitution $\Phi = P(r) \chi(r)$ with some arbitrary functions $W$ and $P$. We would like to find functions $W$, $P$ such that the equation takes the Schr\"odinger form
\begin{equation} \label{eq:kg}
    \chi''
    + \left(
        \omega^2
        - V_\text{eff}
    \right)\chi = 0,
\end{equation}
where prime   denotes the derivative with respect to $\varrho$. To achieve this goal, we must put
\begin{equation} \label{eq:kg_condition}
    W^2 = \frac{v}{a},\qquad
    P = \sqrt{\frac{v}{W R}} = f_2^{\frac{s+d}{\Delta}}
    f_1^{-(1-\sigma)(s+1)/4s}r^{-(s+1)/2}.
\end{equation}
Substituting (\ref{eq:kg_condition}) in Eq. (\ref{eq:dalambertian3}), we will get the effective potential $V_\text{eff}$ for the test scalar field
\begin{align} \label{eq:kg_potential}
    &
    V_\text{eff} =
        {\bm k}^2\frac{a}{b}
        + l(l+s)\frac{a}{w}
        + a \mu^2
        + V_0,
    \\ & \label{eq:kg_v_0}
    V_0 = \left(\ln P\right)'^2 - \left(\ln P\right)''.
\end{align}
The first three terms of $V_\text{eff}$ coincide with the effective potential for geodesics up to the change of constants of motion to the quantum numbers. Note that $L^2$-integrability of the solutions does not depend on the choice of the functions $W$ and $P$, if we choose the correct integration measure in the correspondind  Sturm-Liouville problem: $R/a\, dr$ for (\ref{eq:dalambertian3}), and $d\varrho$ for (\ref{eq:kg}).

Consider the limit $r\to+\infty$. Then, the measure asymptotically tends to $r^{s+1} dr$, so the $L^2$-integrable   scalar field mode should tend to zero faster than $r^{-1-s/2}$. At the same time, the solution of Eq. (\ref{eq:dalambertian3}) has an asymptotic  form
\begin{equation*}
    \Phi \approx r^{-s/2}\left(
        C_1 J_{l+s/2}(\kappa r) + C_2 Y_{l+s/2}(\kappa r)
    \right),\qquad
    \kappa^2 = \omega^2 - {\bm k}^2 - \mu^2.
\end{equation*}
For $\kappa^2 > 0$, the solutions are wave-like and they are not $L^2$-integrable, which is evident from the asymptotic behavior of the oscillation amplitude $r^{-(s+1)/2}$.
The case $\kappa^2=0$ leads to the solution of the form $\Phi \approx C_1 r^l + C_2 r^{-l-s}$, where the only square integrable mode is the second term for $s>2$ or $l>0$.
In the case $\kappa^2 < 0$, one mode exponentially diverges, and another  exponentially decays. So only half of them are $L^2$-integrable.
Summing up, the $L^2$-integrability implies   the inequality $\omega^2 < {\bm k}^2 + \mu^2$ (which can be not strict for $s>2$ or $l>0$).

Consider the solution near the outermost singularity (which is of the Fisher type for $S_G,\,c^2 < 1$).
The asymptotic behavior of the function $W = \sqrt{v/a}$ has been classified in Table \ref{tab:geodesics_1}.
Generally, $W \sim x^m$, where $m$ is some constant, depending on the parameters of the theory. The new coordinate can be expressed in terms of the old one asymptotically as follows $\varrho \sim x^{m+1}$ (or $\rho \sim \ln x$ for $m=-1$) up to a multiplicative constant and an arbitrary additive constant.   The new radial coordinate, which brings the equation to the Schr\"odinger form, strongly depends on whether the geodesics can achieve the singularity with a finite time interval of the distant observer or not.  
Near the singularity the function $P$ behaves as $x^n$ up to a coefficient, where $n$ is some constant.
From Eq. (\ref{eq:kg_v_0}) for $m\neq-1$ follows that $V_0 \approx (\nu^2-1) \varrho^{-2}/4$, where
\begin{equation}\label{eq:kg_typical}
    \nu = \frac{2n+m+1}{m+1}.
\end{equation}

Let us find the form of the solution in the vicinity of the singularity for each metric subfamily.

\paragraph{Solution $S_G$ with $c^2<1$.}
    
The exponent $m=-1+(1+s)(1-\sigma)/2s$ achieves its minimum for $\sigma = 1$  equal to $-1$. The case $\sigma=1$ is similar to the test scalar field case in the background of the regular Schwarzschild or Reissner-Nordstr\"om solutions \cite{Futterman:1988ni, Kehle:2018upl}, so we will not consider it here.
For all other cases we have $m>-1$, so near the singularity $\varrho\to0$.
To avoid terms in $V_\text{eff}$ more singular than $\varrho^{-2}$ from $V_0$, it is necessary to require the following inequality
\begin{equation*}
    \frac{\sigma}{m+1} \geq -2,\qquad \left(\sigma-\frac{1-\sigma}{s}\right)/(m+1) \geq -2.
\end{equation*}
Both conditions strictly hold after the substitution of $m$, so the most singular term of the effective potential is contained in $V_0$.
Function $P$ has an asymptotic $x^{-\frac{m+1}{2}}$, where we find $n=-(m+1)/2$ and $\nu = 0$.
The solution of the Eq. (\ref{eq:kg}) with such an effective potential near $r=r_0$ has the form
\begin{equation} \label{eq:kg_sol_1}
    \Phi \approx P \varrho^{1/2} \left(C_1 + C_2 \ln \varrho \right)
    \approx C'_1 + C'_2 \ln x.
\end{equation}
Both modes are square integrable near the singularity (limit circe). Thus, in this case the singularity is not healed.
The test scalar field is regular if $C'_2=0$ \cite{Bronnikov:2018vbs}.

\paragraph{Solution $S_G$ with $c^2>1$.}
    
With similar calculations we have $m=2(s+d)/\Delta > 0$, $n=-m/2$ and $\varrho \to 0$ near the singularity.
Conditions for the other terms in $V_\text{eff}$ to be less singular than $V_0$, are
\begin{equation*}
    (m+1)\Delta - 2s > 0,\qquad
    (m+1)\Delta - 2(d+s) > 0,
\end{equation*}
and they hold for any parameters.
Substituting the values of $m$ and $n$ in (\ref{eq:kg_typical}), we get $\nu = 1/(m + 1)$.
The solution of Eq. (\ref{eq:kg}) with such an effective potential near the singularity is
\begin{equation} \label{eq:kg_sol_2}
    \Phi \approx P \sqrt{\varrho}\left(
    C_1 \varrho^{+\nu/2} + C_2 \varrho^{-\nu/2}
    \right)\approx
    C_1 x + C_2.
\end{equation}
Both modes are regular and square integrable near the horizon. The singularity is not healed either.
    
\paragraph{Solution $S_E$ with $\alpha^2 \neq \alpha_\text{crit}^2 = 2s^2/(s+d)$.}
    
Such solutions in a theory with $\alpha=0$ are regular extreme Reissner-Nordstr\"om black holes \cite{Dain:2012qw} and will not be considered here, we will assume $\alpha\neq 0$ only.
From the asymptotics of $W$ and $P$ we find 
\begin{equation}\label{eq:as_se_s}
    m=-2s(s+d)/\Delta\neq-1, \qquad
    n=-\frac{m+s+1}{2},
\end{equation}
\begin{equation}
    \varrho \approx \frac{r^{m+1}}{\rho^m (m+1)}.
\end{equation}
Let's keep only the leading terms in the expansion for each term in the effective potential
\begin{equation}\label{eq:Veff_noncrit}
    V_\text{eff} \approx {\bm k}^2 + q \left(\frac{r^{m+1}}{\rho^m (m+1)}\right)^{-2} + \mu^2 \left(\frac{r}{\rho}\right)^{4s^2/\Delta},
\end{equation}
where
\begin{equation}
    q
    =
    \frac{l(l+s)}{(m+1)^2} + \frac{\nu^2-1}{4}
    =
    \left(
        \frac{l+s/2}{m+1}
    \right)^2
    - \frac{1}{4}.
\end{equation}
Solutions will be square integrable near the singularity if $R |\Phi|^2 / a$ decays faster than $r^{-1}$. Taking into consideration Eq. (\ref{eq:limit_2}), an integrable solution $\Phi$ must decay faster than $r^{-s/2 - 1 + 2s(d+s)/\Delta}$ near $r=0$. We will use this condition further to analyze cases $\alpha^2 > \alpha^2_\text{crit}$ and $\alpha^2 < \alpha^2_\text{crit}$ separately.

For $m>-1$ ($\alpha^2>\alpha^2_\text{crit}$) the leading term in Eq. (\ref{eq:kg}) comes from the second term of $V_\text{eff}$, i.e. $q \varrho^{-2}$.
In this case, the solution near the singularity is
\begin{align} \label{eq:kg_sol_3}
    \Phi & \approx
    P\sqrt{\varrho}\Big(
    C_1 \varrho^{\sqrt{q+1/4}}
    +
    C_2 \varrho^{-\sqrt{q+1/4}}
    \Big)
    \approx
    C'_1 r^{l} + C'_2 r^{-l-s}.
\end{align}
The first mode $r^l$ satisfies the square integrability condition if $l > 2 s (d + s)/ \Delta - 1 - s/2$. As we consider the case $\alpha^2 > \alpha^2_\text{crit}$, we can substitute $\alpha^2 = \alpha^2_\text{crit} + x_\alpha$, $x_\alpha > 0$ obtaining a new constraint $l > -1 - s/2 + 2s/(2s+x_\alpha)$. This inequality always holds and the mode $r^l$ is always integrable.
The second mode $r^{-l-s}$ is integrable if $l < -s/2 + x_\alpha / (2s + x_\alpha)$, which is possible only for $l=0$, $s=1$, $\alpha^2 > 2(d+2)/(d+1)$. This mode, however should be excluded from the spectrum on the same ground as  the corresponding  mode in the non-relativistic quantum mechanics (appearance of the delta-function under the action of the full Laplacian). Therefore this case corresponds to the limit point.
    
For the case $m<-1$ ($\alpha^2<\alpha^2_\text{crit}$) with ${\bm k}^2 \neq  \omega^2$, the coordinate $\varrho$ tends to infinity, so the equation for $\chi$ has the form
\begin{equation*}
    \chi'' + \kappa^2 \chi \approx 0,\qquad
    \kappa^2 = \omega^2 - {\bm k}^2.
\end{equation*}
The solution for this equation asymptotically has the form
\begin{equation}\label{eq:kg_sol_4}
    \Phi
    \approx r^n (C_1 \exp(i\kappa \varrho) + C_2 \exp(-i\kappa \varrho)).
\end{equation}
For $\kappa^2>0$ the solution near the singularity oscillates infinitely fast, it diverges due to $n<0$, and  is not integrable.
For $\kappa^2<0$, one of two modes exponentially diverges, and another  exponentially decays, being square integrable near the horizon. However, for $\kappa^2 = 0$ the leading term is $q\varrho^{-2}$. Thus, this case is similar to the previous one $\alpha^2 > \alpha^2_\text{crit}$ with solution (\ref{eq:kg_sol_3}) except the fact that $x_\alpha < 0$ now. As the r.h.s. of the inequality for the square integrability is a monotonic function of $\alpha^2$, we can draw conclusions from the corner cases $\alpha^2 = 0,\,\alpha^2_\text{crit}$. One can find that the mode $r^l$ is integrable except the case $d=1$ with $\alpha^2 \le 2s^2 / (s+1)(s+2)$, and the mode $r^{-l-s}$ is always non-integrable.
    
\paragraph{Solution $S_E$ with $\alpha^2 = \alpha_\text{crit}^2 = 2s^2/(s+d)$.}
    
In this case, we have $m = -1$, $n=-s/2$, $\varrho\approx\rho\ln r \to-\infty$.
Substituting the background solution in Eq. (\ref{eq:kg_v_0}), we get an exact expression for  $V_0$
\begin{equation}\label{eq:v_0_crit}
    V_0 = \frac{1}{4} \left(r^s+\rho ^s\right)^{-\frac{2 (s+1)}{s}} \left(\left(s^2-1\right) r^{2 s}+2 s (s+1) r^s \rho ^s+s^2 \rho ^{2 s}\right).
\end{equation}
In the singular point, the expression (\ref{eq:v_0_crit}) tends to the finite positive value $\left.V_0\right|_{r=0}= (s/2\rho)^2$.
With account for the other terms, the effective potential tends to the value $V_\text{eff} \approx {\bm k}^2 + (l+s/2)^2/\rho^2$.
In this case the equation has the form (\ref{eq:kg_sol_4}) up to the replacement of $\kappa^2 \to \omega^2 -{\bm k}^2 - (l+s/2)^2/\rho^2$ and the dependence $\varrho(r)$.
For $\kappa^2>0$, modes remain singular and non-integrable.
For $\kappa^2<0$ the solution can be simplified
\begin{equation*}
    \Phi
    \approx C_1 r^{-|\kappa| \rho - s/2} + C_2 r^{|\kappa| \rho - s/2}.
\end{equation*}
The first mode is singular and  always non-integrable.
The other mode is regular if $|\kappa|\rho \geq s/2$ (which can be rewritten $\omega^2 \leq {\bm k}^2 + l(l+s)/\rho^2$) and it is always integrable.
In the case $\kappa^2 = 0$, we have to choose the next leading term in the effective potential with the lowest exponent $V_\text{eff} \approx b_1 r^{a_1}$. For the massive scalar field the effective potential asymptotically behaves as $V_\text{eff} \approx \mu^2 (r/\rho)^{a_1}$, where $a_1=2s/(d+s)$, and for the massless case it is $V_\text{eff} \approx s^{-1}\left(s^2 - 2ls - 2l^2\right) \rho^{-s-2} r^s $ with $a_1=s$ (note, the expression $s^2 - 2ls - 2l^2$ cannot be zero for integer $l, s$). Then the solution is
\begin{equation*}
    \chi'' - b_1 \exp\left(a_1 \varrho / \rho\right) \chi \approx 0,
\end{equation*}
\begin{equation*}
    \Phi
    \approx r^{-s/2}\left(
    C_1
    I_0\left(\frac{2 \sqrt{b_1} \rho  }{a_1} r^{a_1/2}\right)
    +
    C_2
    K_0\left(\frac{2 \sqrt{b_1} \rho  }{a_1} r^{a_1/2}\right)
    \right)
    \approx
    C'_1 r^{-s/2}
    +
    C'_2 r^{-s/2}\ln r,
\end{equation*}
where $I_0$, $K_0$ are modified Bessel functions of the second kind.
Solutions we obtained diverge  and they are not integrable for any $C_1$, $C_2$. As there are at most one integrable mode for each set of quantum numbers, the singularity is healed.

The summary of the square integrability and regularity of the modes in the background of the different classes of the solution is given in table \ref{tab:integrability}. The general class is always unhealed since the number of square integrable modes is two. In the case of the Fisher-like singularity ($S_G, c^2 < 1$), one of these modes is integrable. Contrary, in the solution with a singularity of the Reissner-Nordstr\"om type ($S_G, c^2>1$), both modes are regular. In the solution $S_E$, there is at most one square integrable mode, except for one special case. This case is $l=0$, for the solution $S_E, \alpha^2 > 2 \frac{d+2}{d+1} > \alpha_\text{crit}^2$ with $s=1$. However, similarly to quantum mechanics, this mode can be excluded from the spectrum as a solution for the point-like ($\delta$-function distribution) source. If we exclude this mode, all $S_E$ solutions contain a quantum unobservable singularity, otherwise, the singularity is observable only in the special. All square integrable modes are regular except this special $l=0$ case, and one more case with a critical coupling constant $\alpha^2=\alpha_\text{crit}^2$ within the certain interval of the frequency $\frac{l(l + s)}{\rho^2} + {\bm k}^2 < \omega^2 < \left(\frac{l+s/2}{\rho}\right)^2 + {\bm k}^2$.

\begin{table}\caption{The number of square integrable modes and regular modes for each class of solutions and a certain interval of $\omega^2$. If the case contains some special case, this special case is given in the last column and the number of modes for the special case is given in the brackets.  \label{tab:integrability}}
\begin{center}\begin{tabular}{ |c||c|c|c|c| }
 \hline
 Solution &
 Condition on $\omega^2$ &
 \makecell{\# of square \\ int. modes} &
 \makecell{\# of regular \\ modes} &
 Special case
 \\
 \hline\hline
 $S_G,\,c^2<1$
 &
 Any
 &
 2
 &
 1
 &
 \\\hline
 $S_G,\,c^2>1$
 &
 Any
 &
 2
 &
 2
 &
 \\\hline
 $S_E,\,\alpha^2 > \alpha^2_\text{crit}$
 &
 Any
 &
 1 (2)
 &
 1 (1)
 &
 $l=0,\,s=1,\,\alpha^2>2\frac{d+2}{d+1}$
 \\\hline
 \multirow{3}{*}{$S_E,\,\alpha^2 < \alpha^2_\text{crit}$}
 &
 $\omega^2 > {\bm k}^2$
 &
 0
 &
 0
 &
 \\
 &
 $\omega^2 = {\bm k}^2$
 &
 0 (1)
 &
 0 (1)
 &
 $d=1,\,\alpha^2\leq\frac{2s^2}{(s+1)(s+2)}$
 \\
 &
  $\omega^2 < {\bm k}^2$
 &
 1
 &
 1
 &
 \\\hline
 \multirow{4}{*}{$S_E,\,\alpha^2 = \alpha^2_\text{crit}$}
 &
 $\omega^2 - {\bm k}^2 \geq \left(\frac{l+s/2}{\rho}\right)^2$
 &
 0
 &
 0
 &
 \\
 &
 $\frac{l(l + s)}{\rho^2}< \omega^2 - {\bm k}^2 < \left(\frac{l+s/2}{\rho}\right)^2$
 &
 1
 &
 0
 &
 \\
 &
 $\omega^2 - {\bm k}^2 \leq \frac{l(l + s)}{\rho^2}$
 &
 1
 &
 1
 &
 \\
 \hline
\end{tabular}\end{center}
\end{table}

\section{Conclusions}

In this artscle, we have constructed charged $p$-branes with primary scalar charge in Einstein gravity with dilaton and antisymmetric forms using Harrison transformations  adapted for branes  applied to an extended Fisher solution. The obtained branes, spherically symmetric in the transverse space,   generically  have naked singularities, except for those obtained from the regular subfamily of the seed solutions  and Harrison transformation parameter $c^2<1$. 

Using some limiting procedure in the space of parameters we also found a special solution,   denoted as $S_E$,    satisfying a constraint on the physical charges which  reduces to the ``no-force'' condition in EMD theory with $\alpha^2=3$. Exploring supersymmetry of the full new family of solutions corresponding to NS5 branes we found that supersymmetry  holds only in the limiting case $S_E$, which coincides with the standard BPS NS5-branes.

We investigated the geodesic motion in the vicinity of the outermost singularity of the obtained solutions.
In the general case with $c^2<1$ with Fisher singularity, the parameter $\sigma$ plays the key role in the behavior of the geodesics.
In the general case with $c^2>1$, the behavior does not differ appreciably for different theory parameters and charges of a solution.
The behavior of motion near the singularity for $S_E$ family depends on the relation between the coupling constant $\alpha$ and its critical value $\alpha^2_\text{crit} = 2s^2/(s+d)$.
If $\alpha^2\leq\alpha^2_\text{crit}$, geodesics near the singularity behave similarly to the motion near a black hole surface: the effective potential is always bounded, a particle achieves the singularity with an infinite time with respect to a distant observer.
Otherwise, if $\alpha^2>\alpha^2_\text{crit}$ the effective potential for geodesics with non-zero angular moment diverges, but radial geodesics achieve the singularity with a finite time of a distant observer.

We considered a massive test scalar field in the background of the obtained solutions.
By transforming the dynamical variable and the radial coordinate, one can represent the Klein-Gordon equation in the form of the Schr\"odinger equation with some effective potential. We investigated the possibility to heal the singularity in a quantum sense, based on the analysis of square integrability of the test field.
For the solution family $S_E$, it was shown that the behavior of the test scalar field depends on whether the coupling constant $\alpha$ get over the critical value $\alpha_\text{crit}$ or not. For $S_E$, at most one mode is square integrable near the singularity (and so the singularity is healed), except the case $s=1$, $\alpha^2 > 2(d+2)/(d+1) > \alpha^2_\text{crit}$ for the mode $l=0$.   But this latter mode should be excluded from the spectrum of perturbations as a solution for a point-like source, similarly to the non-relativistic quantum mechanics. 

For a general family with arbitrary $c^2$, all modes are square integrable near the singularity. Thus, the choice of boundary conditions remains an important issue of the stability of the solution with respect to test disturbances of the scalar field in the general case. 

Finally, in the case of codimension three, we constructed the branes equipped with Zipoy-Voorhees oblateness parameter, which do not have spherical symmetry in transverse space. These solutions do not seem to be known before.

\begin{acknowledgments}
This work was partially supported by the the Russian Foundation for Basic Research, project 20-52-18012Bulg-a, the Scientific and Educational School of Fundamental and Applied Space Research of the Moscow State University, and the Russian Government Program of Competitive Growth of the Kazan Federal University. I.B. is also grateful to the Foundation for the Advancement of Theoretical Physics and Mathematics ``BASIS'' for support and K.V. Kobialko for useful discussions.

\end{acknowledgments}

\end{document}